\documentclass[%
 reprint,
superscriptaddress,
preprintnumbers,
nofootinbib,
 amsmath,amssymb,
 aps,
]{revtex4-1}

\usepackage[colorlinks]{hyperref}
\usepackage{graphicx} 
\usepackage{dcolumn}
\usepackage{bm}
\usepackage{color}
\usepackage{slashed}
\usepackage[dvipsnames]{xcolor}
\usepackage{placeins}
\usepackage{slashed}
\usepackage{multirow}
\usepackage{subfigure}
\usepackage{tabularx}
\usepackage{mathtools}
\usepackage{floatrow}
\usepackage{blindtext}
\usepackage{soul}
\usepackage{slashed}
\usepackage{amsmath}
\usepackage{braket}
\usepackage{comment}

\hypersetup{
  colorlinks=true,
  linkcolor=red,
  citecolor=blue,
  urlcolor=blue,
}

\definecolor{coral}{RGB}{255,127,80}
\definecolor{indigo}{RGB}{75,0,130}
\definecolor{red}{rgb}{0.9, 0,0}
\definecolor{cerulean}{rgb}{0., 0.62,0.9}
\definecolor{navy}{rgb}{0.05, 0.05,0.8}

\newfloatcommand{capbtabbox}{table}[][\FBwidth]

\def\NANO{{\sc NANOGrav}}
\def\EPTA{{\sc EPTA}}
\def\PPTA{{\sc PPTA}}

\def\IPTA{{\sc IPTA30}}

\begin{document}
\title{Bounds on Ultralight Dark Matter from NANOGrav}

\author{Mohammad Aghaie}
\affiliation{Dipartimento di Fisica E. Fermi, Universit\`a di Pisa, Largo B. Pontecorvo 3, I-56127 Pisa, Italy }
\affiliation{INFN, Sezione di Pisa, Largo Bruno Pontecorvo 3, I-56127 Pisa, Italy}

\author{Giovanni Armando}
\affiliation{Dipartimento di Fisica E. Fermi, Universit\`a di Pisa, Largo B. Pontecorvo 3, I-56127 Pisa, Italy }
\affiliation{INFN, Sezione di Pisa, Largo Bruno Pontecorvo 3, I-56127 Pisa, Italy}

\author{Alessandro Dondarini}
\affiliation{Dipartimento di Fisica E. Fermi, Universit\`a di Pisa, Largo B. Pontecorvo 3, I-56127 Pisa, Italy }
\affiliation{INFN, Sezione di Pisa, Largo Bruno Pontecorvo 3, I-56127 Pisa, Italy}

\author{Paolo Panci}
\affiliation{Dipartimento di Fisica E. Fermi, Universit\`a di Pisa, Largo B. Pontecorvo 3, I-56127 Pisa, Italy }
\affiliation{INFN, Sezione di Pisa, Largo Bruno Pontecorvo 3, I-56127 Pisa, Italy}


\begin{abstract}
 The compelling evidence for the detection of the stochastic gravitational wave background by \NANO\ imposes constraints on the mass of compact cores of ultra-light dark matter, also known as \lq\lq solitons", surrounding supermassive black holes found at the centers of large galaxies. The strong dynamical friction between the rotating black holes and the solitons competes with gravitational emission, resulting in a suppression of the characteristic strain in the nHz frequency range. Our findings robustly rule out ultralight dark matter particles with masses ranging from $1.3\times 10^{-21}$~eV to $1.4\times 10^{-20}$~eV condensing into solitons around supermassive black holes.
\end{abstract}

\maketitle

\section{Introduction}\label{Intro}
The detection of the stochastic gravitational wave background (SGWB) by pulsar timing array (PTA) observations would be monumental because it could provide direct evidence of the existence of SuperMassive Black Hole (SMBH) binaries and further confirm the predictions of general relativity. The major PTA collaborations currently looking for SGWB are: the European Pulsar Timing Array (\EPTA) ~\cite{Antoniadis:2023rey,Antoniadis:2023zhi}, the
Parkes Pulsar Timing Array (\PPTA)~\cite{Reardon:2023gzh, Reardon:2023zen, Zic:2023gta}, \NANO~\cite{NANOGrav:2023hfp}, the MeerKAT Pulsar Timing Array~\cite{Miles:2022lkg}, the Indian Pulsar Timing Array (InPTA)~\cite{EPTA:2023xxk}, the Chinese Pulsar Timing Array (CPTA)~\cite{Xu:2023wog}, and the International Pulsar timing array(IPTA) that combines the data from several PTA experiments~\cite{InternationalPulsarTimingArray:2023mzf}. On a more specific level, PTAs use the precise timing measurements of highly stable millisecond pulsars to detect modulations in the propagation of light caused by Gravitational Waves (GWs). Such modulations are correlated with the angular separation of the pulsars as noted by Hellings and Down~\cite{Hellings:1983fr} and have recently been confirmed by the  \NANO\ collaboration~\cite{NANOGrav:2023hfp}.  

\smallskip
The  measured SGWB is approximately consistent with the gravitational wave background (GWB) produced by inspiralling SMBH binaries. Importantly, SMBHs with masses ranging from $10^5$ to $10^{10}$ $M_{\odot}$ reside at the centers of galaxies. Throughout the hierarchical process of structure coalescing, these galaxies might merge and give rise to SMBH binaries, at the late stages of the merger event \cite{Begelman:1980vb, McConnell:2012hz}. Eventually, towards the end of their evolution, SMBH binaries produce gravitational waves in the nHz frequency band that is detected by PTAs. If the SMBH binary evolution is driven solely by the loss of orbital energy to gravitational waves, then the SGWB is described by a \textit{characteristic strain} $h_c$ with power-law behaviour in frequency $h_c(f)\propto f^{-2/3}$~\cite{Phinney:2001di}, although a realistic modeling of the astrophysical environment surrounding the binaries predicts slightly more attenuated power-laws~\cite{NANOGrav:2023hfp}. Moreover, potential systematic effects in pulsar noise modeling could lead to alterations in the spectrum (see e.g.~\cite{Reardon:2023gzh}).

\smallskip
On top of the purely astrophysical effects, the modification to the standard power-law index induced by new physics around the binaries can be foreseen. The \NANO\ collaboration has considered several Beyond Standard Model scenarios that can source or contribute to the SGWB~\cite{NANOGrav:2023hvm}. For example, cosmological sources like inflation and primordial fluctuations~\cite{Vagnozzi:2020gtf, Benetti:2021uea, Vagnozzi:2023lwo, Franciolini:2023pbf, Franciolini:2023wjm, Inomata:2023zup, Ebadi:2023xhq, Cai:2023dls, Wang:2023ost, Gouttenoire:2023nzr, Liu:2023ymk, Abe:2023yrw, Unal:2023srk, Yi:2023mbm, Firouzjahi:2023lzg, Salvio:2023ynn, You:2023rmn, Bari:2023rcw, Ye:2023xyr, Balaji:2023ehk, HosseiniMansoori:2023mqh, Cheung:2023ihl, Das:2023nmm, Jin:2023wri, Bousder:2023ida, Zhao:2023joc, Yi:2023tdk, Ben-Dayan:2023lwd, Jiang:2023gfe, Liu:2023pau, Frosina:2023nxu, Bhaumik:2023wmw}, cosmic strings~\cite{Ellis:2023tsl, Kitajima:2023vre, Wang:2023len, Lazarides:2023ksx, Eichhorn:2023gat, Chowdhury:2023opo, Servant:2023mwt, Yamada:2023thl, Antusch:2023zjk, Ge:2023rce, Basilakos:2023xof}, first-order phase transitions~\cite{Xue:2021gyq, NANOGrav:2021flc, Addazi:2023jvg, Bai:2023cqj, Megias:2023kiy, Han:2023olf, Zu:2023olm, Ghosh:2023aum, Xiao:2023dbb, Li:2023bxy, DiBari:2023upq, Cruz:2023lnq, Gouttenoire:2023bqy, Ahmadvand:2023lpp, An:2023jxf, Wang:2023bbc}, topological defects~\cite{Blasi:2020mfx, Ellis:2020ena, Kitajima:2023cek, Guo:2023hyp, Blasi:2023sej, Gouttenoire:2023ftk, Barman:2023fad, Lu:2023mcz, Li:2023tdx, Du:2023qvj, Babichev:2023pbf, Gelmini:2023kvo, Zhang:2023nrs}, and “audible” axions~\cite{Figueroa:2023zhu, Geller:2023shn} can improve the fit to the observed data, although this mainly stems from insufficient statistics. Another interesting possibility among the new physics scenarios that may be involved in the production of a SGWB  is provided by the dynamical friction of the rotating BHs in a dense Dark Matter (DM) environment (see~\cite{Ghoshal:2023fhh} for a preliminary study), if this dominates over the one induced by baryons. 
  
\smallskip
In this paper we demonstrate that the measurement of a SGWB in the nHz frequency range provides a unique opportunity  to probe some properties of  Ultra-Light Dark Matter (ULDM).  ULDM is an appealing alternative to cold DM  because in the mass window $10^{-22}\,\text{eV} \lesssim m \lesssim 10^{-20}\,\text{eV}$ it has a de Broglie wavelength of the order of few kpc and therefore exhibits a wave-like behavior at small galactic scales. The wave-like nature of ULDM particles introduces quantum pressure effects that can counteract gravitational collapse and prevent the formation of cuspy density profiles alleviating some of small-scale puzzles facing the cold DM paradigm \cite{Hu:2000ke, DelPopolo:2016emo, Hui:2016ltb}. 
Furthermore, ULDM particles are expected to condense into a dense core, generally known as ``soliton"~(see e.g.~\cite{Hui:2016ltb, Bar:2019pnz}). The problem of a massive scalar field, minimally-coupled in the strong gravity regime of a black hole (BH) (including the phenomenon of superradiance~\cite{Brito:2015oca}), has been extensively studied in the literature (see, for example,~\cite{Arvanitaki:2009fg}). Other papers~\cite{Hui:2016ltb, Bar:2018acw} have explored the interaction between BH and ULDM on galactic scales, using the Newtonian approximation. Similarly to~\cite{Bar:2018acw}, our approach focuses on the intermediate regime, where a Newtonian analysis is applicable, but the dynamics are predominantly governed by the SMBH.

\smallskip
The ULDM soliton could be detected given detailed knowledge of the late stages of the SMBH binaries merger event. Indeed, as the separation of the SMBH binaries falls below the soliton radius, both BHs will experience a wind of ULDM particles. This phenomenon, known as \textit{dynamical friction}, introduces a new energy loss mechanism that can compete with gravitational wave emission in the nHz range. Therefore, the evolution of SMBH binaries is modified in the presence of solitons. In the end, the current PTA data are able to constrain such a modification in the spectral shape of the SGWB, setting limits on the ULDM particle and soliton masses.

\smallskip
This paper is organized as follows. Sec.~\ref{Sec:ULDM} sets the stage for our investigation, introducing basic properties of the ULDM soliton, explaining with ballpark numbers how a dense core of ULDM particles can dramatically affect the emission of GWs in the nHz frequency range. In addition, we also discuss the issue of the depletion of the soliton and which constraints DM numerical simulations put on the mass of the soliton. In Sec.~\ref{Sec:strain} we delve into the calculations of the characteristic strain while considering the impact of dynamical friction originating from ULDM particles. In Sec.~\ref{Sec:results} we present our results in light of the recent \NANO\ dataset and the comparison with existing bounds. In Sec.~\ref{Sec:conclusions} we draw our conclusions. 

\section{Ultralight Dark Matter}\label{Sec:ULDM}
\subsection{The Soliton shape}

The phenomenology of ULDM gives rise to a dense macroscopic core in the galactic center, called a soliton. The soliton corresponds to a quasi-stationary minimum energy solution of the equations of motion of the bosonic ultralight real scalar field $\phi$ with mass $m$, minimally-coupled to gravity~\cite{Hui:2016ltb, Bar:2019pnz}. Here we are interested in the soliton solution in the presence of a SMBH. Following~\cite{Bar:2019pnz}, we decompose $\phi$ as 
\begin{equation}\label{phi}
\phi(r, t) = \frac{e^{-im(1+\gamma)t}}{\sqrt{8\pi G_{\rm N}}}\chi(r) + \text{c.c.}\,.
\end{equation}
where $G_{\rm N}$ is the Newton's constant and $\gamma$ is an eigenvalue of the problem. 

\smallskip
Working in the non-relativistic regime and assuming spherical symmetry, the  Klein-Gordon and Poisson equations are reduced to
\begin{equation}\label{eq:KG}
\begin{split}
     &\partial_r^2(r\chi )= 2r\left(m^2\Phi -\frac{A m}{r} - m^2 \gamma\right)\chi ,\\
     &\partial _r^2(r\Phi )= m^2 r\chi^2, 
     \end{split}
\end{equation}
where $A = G_{\rm N} M_{\bullet} m$, $M_{\bullet}$ is the BH mass and $\Phi$ is the Newtonian gravitational potential. In the ULDM mass range considered in our analysis and with respect to the SMBH peak mass favored by the \NANO\ data~\cite{NANOGrav:2023hfp}, the parameter $A$ is significantly less than one, and as a result, the Newtonian approximation is well-justified. Within this regime, relativistic phenomen\ae\ are negligible.  

Solving the equations above numerically is not particularly difficult and one can find a careful treatment in~\cite{Bar:2019pnz, Hui:2016ltb, Marsh:2015wka, Davies:2019wgi}. In this study, we focus on the regime where the soliton profile is primarily dominated by the BH gravity: there the soliton-BH mass ratio  $\varepsilon=M_{\rm sol}/M_{\bullet} \lesssim 0.2$ as one can infer from  Fig.~7 of~\cite{Bar:2019pnz}. This is because self-gravitating solitons with a mass equal to or greater than that of the SMBH significantly affect the rotational curves of the host galaxy~\cite{Bar:2018acw}. More specifically, the density  is given by
\begin{equation}\label{eq:rho}
\rho(r;A) \approx \frac{m^2}{ \pi G_{\rm N}} \lambda^4  \, e^{-2 A m \, r} \equiv \frac{m^2 \varepsilon}{ \pi G_{\rm N}} A^4  \, e^{-2 A m \, r} \ ,
\end{equation}
where $\lambda$ is a continuous positive parameter choosen as the value of the field at the origin~\cite{Bar:2019pnz}. 

As a consequence, an estimate for the soliton core density $\rho_0\equiv\rho(r=0)$ writes 
\begin{equation} \label{eq:coredensity}
      \rho_0[M_\odot \text{pc}^{-3}] \simeq 5\times 10^4  \varepsilon \, M_{\bullet}^4[10^8 M_{\odot}]\, m^6[10^{-21} \, \text{eV}] \ ,
      \end{equation}
 where $M_{\bullet}[10^8 M_{\odot}]$, $m[10^{-21} \, \text{eV}]$ are the BH and point particle masses in units of $10^8 M_{\odot}$ and $10^{-21} \, \text{eV}$ respectively. Similarly, the soliton radius\footnote{We define the soliton radius as the distance by which the density is half of the soliton central density.} is independent on the soliton mass and can be estimated from Eq.~\eqref{eq:rho} as $r_{\rm sol}\sim (\ln{2})/2A m$, which leads to
 \begin{equation}\label{eq:coreradius}
      r_\text{sol}[\text{pc}] \simeq \frac{2.9}{M_{\bullet}[10^8 M_{\odot}]\, m^2[10^{-21} \, \text{eV}]}  \ .
 \end{equation}

 \begin{figure}[t!]
\begin{center}
	\includegraphics[width=.95\textwidth]{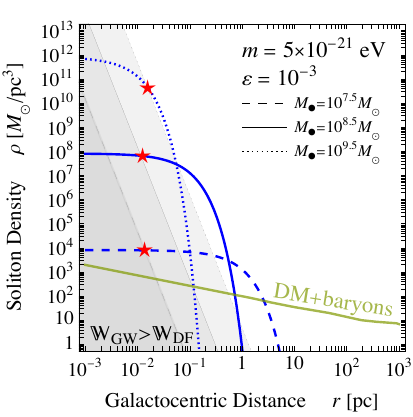}
	\caption{Soliton density as a function of the distance from the SMBH center keeping fixed  $m=5\times 10^{-21}$ eV and $\varepsilon = 10^{-3}$. The blue lines refer to different choices of the SMBH mass, while the green line represents the typical density of the surrounding SMBH environment. The gray shaded areas indicate the region of the parameter space where the energy losses due to ULDM dynamical friction are subdominant with respect to the ones in GWs. The red stars denote the critical radius above which the energy losses due to dynamical friction dominate if smaller than $r_{\rm sol}$.}\label{fig:SolitonDensity}
\end{center}
\end{figure}

Fig.~\ref{fig:SolitonDensity} shows the soliton density as a function of the distance from the central SMBH considering $m=5\times 10^{-21}$ eV and a soliton-BH mass ratio $\varepsilon = 10^{-3}$. The blue lines are obtained by considering respectively $M_{\bullet}=10^{7.5}M_\odot$ (dashed line), $M_{\bullet}=10^{8.5}M_\odot$ (solid line) and $M_{\bullet}=10^{9.5}M_\odot$ (dotted line). For reference we also report  in green the density of  the surrounding SMBH environment composed by stars, ordinary DM and gas~\cite{Kelley:2016gse}. For a given value of the SMBH mass, the gray shaded areas indicate the region of the parameter space where the energy losses due to ULDM dynamical friction are subdominant with respect to the ones in GWs. As one can see, below a few pc the density around the SMBH is totally dominated by the ULDM soliton. As a consequence, unlike the standard cold DM picture, the effects induced by the matter surrounding the binary plays a negligible role and the main impact on the GW emission is fully controlled by the soliton core through  ULDM dynamical friction as we will discuss in more details in Sec.~\ref{dynfricsol}.

\subsection{How much mass into the soliton}\label{Sec:SolMass}
For DM-only numerical simulations, the mass of the soliton $M_{\rm sol}$ is related to the galactic halo mass $M_h[10^{12} M_{\odot}]$ in units of $10^{12} M_{\odot}$ via the relation~\cite{Schive:2014dra,Schive:2014hza} 
\begin{equation}\label{eq:Msol_sim}
M_{\rm sol}^{\rm NS}[M_\odot ]\simeq \frac{1.4 \times 10^8}{m[10^{-21} \, \text{eV}] } M_h^{\frac13} [10^{12} M_{\odot}]  \ ,
\end{equation}
which is equivalent to state that the  kinetic energy per particle of ULDM  is the same in the host galactic halo and in the central soliton core.  We will use Eq.~\eqref{eq:Msol_sim} as an upper bound with which the \NANO\ constraint in the $(m,\varepsilon)$ plane can be superimposed.   In this case, it is indeed  plausible that  the soliton may form early and precede the formation of the SMBH. Hence, there is no reason to expect that SMBH formation would quench the soliton yielding a mass that agrees with the numerical simulation in Eq.~\eqref{eq:Msol_sim}.

\smallskip
Nevertheless, it is worth stressing that the N-body simulations obtain the result in Eq.~\eqref{eq:Msol_sim} with two important caveats: $i)$ they are performed by fixing $m=10^{-22}$ eV and $M_{\rm h}= (10^9 - 5\times 10^{11})M_{\odot}$; and $ii)$ they do not take into account the possibility that the SMBH preceded
the formation of the soliton. As a consequence it is important to provide an estimate on $M_{\rm sol}$ in the limit where the soliton profile is completely controlled by the SMBH-gravity. Following Ref.~\cite{Bar:2019pnz}, we start from the theoretical prediction of the mass of a self-gravitated soliton ($A\to 0$) given by $M_{\rm sol}^{\rm NS}=2.06 \, \lambda / G_{\rm N}m$ (see Appendix~A of~\cite{Bar:2019pnz}) and we compare it to Eq.~\eqref{eq:Msol_sim} to extract a  physical insight of the parameter  $\lambda$  in terms of $m$ and $M_h$. Then, by integrating Eq.~\eqref{eq:rho} and assuming that the parameter $\lambda$ is of the same order of the one for a self-gravitating soliton, $M_{\rm sol}$  in the SMBH dominance writes
\begin{equation}\label{eq:Msol_BH}
M_{\rm sol}^{\rm BH}[M_\odot ]\simeq \frac{5.4 \times 10^6}{m^4[10^{-21} \, \text{eV}] \, M_{\bullet}^3[10^8 M_{\odot}]} M_h^{\frac43} [10^{12} M_{\odot}]  \ .
\end{equation}
This estimate is smaller than the one in Eq.~\eqref{eq:Msol_sim} since the soliton formation is halted by the dynamical heating due to the SMBH when $A^2/2$ is larger than the  kinetic energy per particle of ULDM  in the host galactic halo. Hence, we will use Eq.~\eqref{eq:Msol_BH} as a lower bound with which the \NANO\ constraint in the $(m,\varepsilon)$ plane can be superimposed.

\subsection{Absorption of soliton by SMBH}\label{Sec:abstime}

 A solitonic core around a BH can be depleted in the BH dominant case. The computation of the soliton lifetime has been considered by many authors~\cite{Bar:2019pnz, Hui:2016ltb, Barranco:2017aes, Urena-Lopez:2002nup, Cardoso:2022nzc}. In particular, Ref.~\cite{Barranco:2017aes} gives an estimate of the absorption time by considering the soliton as a quasi-bound solution of the Klein-Gordon equation in a Schwarzschild background and focuses on 0-angular momentum states only, $l=0$. In this regime, the absorption time reads 
 \begin{equation}\label{eq:abstime}
     \tau_{\rm abs}[\text{Gyr}]\simeq  5.6\times10^3 M_{\bullet}^{-5}[10^8 M_{\odot}] m^{-6}[10^{-21}\,\rm eV] \ ,
 \end{equation}
 Notice that the absorption time is extremely sensitive to the BH and ULDM particle masses but it is independent on the mass of the soliton. However, we stress that Eq.~\eqref{eq:abstime} does not take into account the effects of the gravitational backreaction of the surrounding ULDM halo and the possible re-condensation of the soliton. These effects may become relevant when the halo is much more massive than the central SMBH and can substantially alter the absorption time. For example, the problem related to the gravitational backreaction issue has been partially tackled by Ref.~\cite{Barranco:2017aes} for very specific cases, which however, do not include the regime of our analysis. A careful study of the characteristic absorption time in a realistic astrophysical environment may give more robust estimates on the depletion of the soliton. For this reason, we only consider Eq.~\eqref{eq:abstime} as a benchmark timescale keeping well in mind that possible stabilizing effects could make the soliton cores even longer-lived. 
 
\subsection{Dynamical friction}\label{dynfricsol}
Another feature of ULDM that is relevant for our study on the GW emission is \textit{dynamical friction}. This contribution is computed by taking into account the gravitational feedback on a test mass $m_{\rm cl}$ (in our relevant case the BH, $m_{\rm cl}=M_{\bullet}$) moving in a background of ULDM particles with density $\rho$ \cite{Hui:2016ltb, Chandrasekhar:1943ys, Lancaster:2019mde, Vicente:2022ivh, Traykova:2021dua, Traykova:2023qyv}. In particular, Ref.~\cite{Hui:2016ltb} derives dynamical friction in the test mass rest frame. In this way, the problem can be traced back to Coulomb scattering and the application of Gauss's theorem on a sphere of radius $r$ centered on the test object 
leads to the following classical friction force 
\begin{equation}\label{dynfric}
     F_{\rm DF}=\frac{4\pi G_{\rm N}^2 m_{\rm cl}^2\rho}{v^2}C_{\rm cl}(\Lambda) \ .
\end{equation}
Here, $v$ is the velocity of ULDM particles approaching the test mass, and $C_{\rm cl}(\Lambda)=\log(2\Lambda)-1+(1/\Lambda)\log(2\Lambda)$ is a function of the cutoff $\Lambda=v^2r/(G_{\rm N} m_{\rm cl})$ which originates from the infra-red divergence of the Coulomb scattering. In the following $r$ is taken to be the binary separation as elaborated further in Sec.~\ref{Sec:strain}.

It is worth mentioning that Eq.~(\ref{dynfric}) was originally derived for an object moving in a straight line, and we are here extending its application to an object in circular orbit (see~\cite{Annulli:2020lyc, Buehler:2022tmr} for analysis of circular motion). Furthermore, we emphasize that in our analysis, the ULDM is non-relativistic and far from the strong gravity regime ($A\ll 1$). Consequently, the ULDM gravitational impact parameter is comparable to or larger than its de Broglie wavelength. In this limit, the classical expression for dynamical friction is well justified.

\section{GW Spectrum}\label{Sec:strain}
In this section, we study the GW frequency spectrum emitted by the merging of SMBH binaries in the presence of ULDM friction. 
In the case of light BH binaries, it is not immediately clear if complete merging can occur due to quantum fluctuations stored in the ULDM halos of their host galaxies. These fluctuations might hinder the inspiralling process, as first pointed out in~\cite{Hui:2016ltb} and more recently in~\cite{Bar-Or:2018pxz}. These studies provide an expression for the stalling radius $r_{\text{stall}}$ within the framework of self-gravitating ULDM. However, while this consideration may also apply in scenarios where the soliton profile is dominated by the SMBH gravity\footnote{A similar extrapolation from the self-gravitating ULDM regime to BH dominance was undertaken in Sec.~\ref{Sec:SolMass}  to estimate the soliton mass, assuming consistency of the parameter $\lambda$ across both regimes.}, the mergers analyzed in our work involve such massive objects ($M_\bullet$ bigger than $10^7 M_\odot$) that stalling is unlikely to occur (see Fig.~4 of~\cite{Bar-Or:2018pxz}). 

\smallskip
We conveniently describe the frequency spectrum of GWs in terms of the characteristic strain $h_c$. This is parametrized as a power-law and is related to the GW density $\Omega_{\rm GW}$ as
\begin{equation}
    h_c^2(f)= \frac{3 H_0^2}{2\pi^2 f^2}\Omega_{\rm GW}(f)=\left[A_{\rm GW}\left(\frac{f}{1\,\rm yr^{-1}}\right)^{\beta}\right]^2,
\end{equation}
 where $H_0$ is the present value of the Hubble parameter, $f$ is the observed GW frequency while $A_{\rm GW}$ and $\beta$ are the GW amplitude and power index respectively. For \lq\lq pure" GW emission from SMBH binaries, $\beta_{\rm GW}=-2/3$. 
 
 \smallskip
 At the practical level, the strain $h_c$ is computed by taking into account the whole SMBH binary density population convoluted with the emitted energy in GW of the single merger event \cite{Phinney:2001di}
 \begin{equation}\label{eq:strain}
     h_c^2(f)=\frac{3 H_0^2}{2\pi^2 \rho_c f^2}\int \text{d}z\,\text{d}\mathbf{X}\frac{\text{d} n_s}{\text{d}z \, \text{d}\mathbf{X}}\frac{f_s}{1+z}\frac{\text{d}E_{\rm GW}}{\text{d}f_s}\bigg|_{\mathbf{X}} \ .
 \end{equation}
 Here, $z$ denotes the redshift, $n_s$ is the comoving number density of the sources, $f_s=f (1+z)$ is the frequency in the frame of the source and $\rho_c$ is the critical density of the Universe. The variable $\mathbf{X}$ denotes collectively the parameters that characterize the single binary source, such as the two BH masses. The emitted GW energy per unit frequency $\text{d}E_{\rm GW}/\text{d}f_s$ can be easily computed from classical mechanics considerations. In the limit where the soliton solution is dominated by the BH gravity, the mass of the soliton is negligible compared to the ones of the BHs in the binary. Hence, following~\cite{Dror:2021wrl} and taking into account the energy losses due to the ULDM dynamical friction and the emission of GWs, we get
 \begin{align}\label{eq:dEdfsbare}
     \frac{\text{d} E_{\rm GW}}{\text{d}f_s}=\frac{\mu}{3}\left[\pi G_{\rm N}(M_{\bullet, 1}+M_{\bullet,2})\right]^{\frac{2}{3}} \hspace{-.1cm} f_s^{-\frac{1}{3}} \hspace{-.1cm} \frac{\mathcal W_{\rm GW}}{\mathcal W_{\rm GW}+\mathcal W_{\rm DF}} \ ,
 \end{align}
where $M_{\bullet,1}$ and $M_{\bullet,2}$ are the masses of the two SMBHs in the binary, $\mu=M_{\bullet,1}M_{\bullet,2}/(M_{\bullet,1}+M_{\bullet,2})$ is the reduced mass of the system, and $\mathcal W_{\rm GW}$ and $\mathcal W_{\rm DF}$ are the dissipated power in GW emission and ULDM friction, respectively. 

The first contribution is given by (see e.g. Ref. \cite{Maggiore:2007ulw}) 
 \begin{equation}\label{WGW}
     \mathcal W_{\rm GW}=\frac{32}{5} G_{\rm N} \mu^2 \omega^6 r^4,
 \end{equation}
 where the orbital angular velocity of the system $\omega=\pi f_s$ is related to the binary separation $r$ through the standard Kepler's law
 $ \omega= \sqrt{G_{\rm N}(M_{\bullet,1}+M_{\bullet,2})/r^3}$. 
 
 The $\mathcal W_{\rm DF}$ contribution in Eq.~\eqref{eq:dEdfsbare} is on the other hand obtained from Eq.~\eqref{dynfric} in the binary rest frame: namely by replacing $m_{\rm cl}\to \mu$ and $v$ with the relative velocity  $v_{\rm rel}$ between the approaching SMBH and the ULDM particles into the soliton. In the limit where the BH gravity dominates, the velocity of the ULDM particles is equal to $A$ which has to be compared with the orbital velocity $\omega r$ of the SMBH. In the nHz frequency range and for $m \approx  10^{-21}$ eV the velocity of the 
 SMBH with mass smaller than $\approx 10^{9.5} M_\odot$ is always larger than $A$  and therefore $v_{\rm rel} \simeq \omega r$ in the majority of the parameter space considered in our analysis. With this assumption 
  \begin{equation}\label{WDM}
      \mathcal W_{\rm DF}=F_{\rm DF} v_{\rm rel} = \frac{4\pi G_{\rm N}^2\mu^2\rho}{\omega r }C_{\rm cl}(\tilde\Lambda) \ ,
  \end{equation}
with $\rho$ given by Eq.~\eqref{eq:rho} and the cutoff $\tilde\Lambda$ is obtained by using Kepler's law: $\tilde\Lambda = M_{\bullet,1}/M_{\bullet,2}(1+M_{\bullet,2}/M_{\bullet,1})^2$. 
From now on we assume $M_{\bullet,1}\equiv M_\bullet>M_{\bullet,2}$ and rewrite the main quantities in terms of the ratio $q_{\bullet}=M_{\bullet,2}/M_{\bullet,1}<1$.

\smallskip
It is now interesting to compare the dissipative powers  $\mathcal W_{\rm GW}$ and $\mathcal W_{\rm DF}$  to give an estimate of the critical frequency where depletion of the GW signal due to dynamical friction is expected. The gray shaded areas in Fig.~\ref{fig:SolitonDensity}  show the region in the $(r,\rho)$ plane where the energy losses due to ULDM dynamical friction are subdominant with respect to the ones in GWs. On a more specific level, if the critical radius individuated by the interception between the blue lines and the corresponding boundary of the gray regions is smaller than $r_{\rm sol}$, the energy losses due to dynamical friction  completely change the  behaviour of the standard SGWB.   A rough estimate of the critical radius is obtained by imposing $\mathcal W_{\rm DF}/\mathcal W_{\rm GW} = 1$ and $\rho=\rho_0$. One gets 
\begin{equation}\label{eq:rc}
r_c=\sqrt[11]{\frac{64 \left(1+q_{\bullet}\right)^7}{25 \, C_{\rm cl}^2(\tilde\Lambda)}\left(\frac{1}{G_{\rm N} m^{12} M_\bullet \varepsilon^2}\right)} \ ,
\end{equation}
which barely depends on $M_{\bullet}$ and $\varepsilon$. For example, considering the benchmarks in Fig.~\ref{fig:SolitonDensity}, the critical radius is of the order of $10^{-2}$ pc (denoted as red stars in the figure) which corresponds to a critical frequency around a few nHz. For a soliton surrounding a $10^{9.5}M_\odot$ BH, $r_{\rm sol}<r_{\rm c}$ and therefore we do not expect any contribution from dynamical friction. For the other two benchmarks,  $r_{\rm sol} > r_{\rm c}$ and therefore a relevant depletion of the GW signal in the nHz frequency band is foreseen. This will give us a unique opportunity to probe the effects of ULDM condensation around SMBHs by using current PTAs experiments.

\begin{figure*}[t!]
\begin{center}
	\includegraphics[width=.495\textwidth]{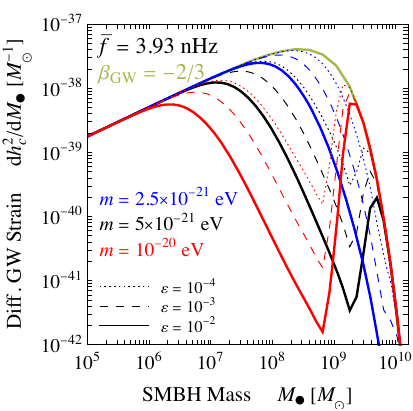}\hfill
	\includegraphics[width=.48\textwidth]{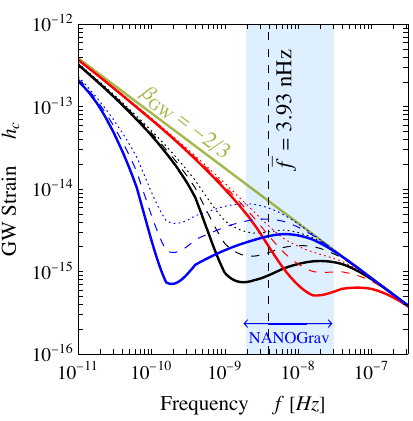}\hfill
	
	\caption{{\it Left-Panel:} Differential GW strain squared as a function of $M_\bullet$ computed keeping  the frequency $f=\bar{f}$ fixed. {\it Right-Panel:} GW strain as a function of $f$. The blue region refers to the \NANO\  frequency band, while the vertical dashed line indicates the frequency $f=\bar{f}$. In all plots three benchmark values of the ULDM particles $m$ are denoted as colored lines, while the benchmark BH-soliton mass ratios $\varepsilon$ are reported with different dashing. The pure GW strains which only come from SMBH mergers are instead shown in green. \label{fig:strains}}
\end{center}
\end{figure*}

\smallskip
Inserting the loss powers $\mathcal W_{\rm GW}$ and  $\mathcal W_{\rm DF}$  in Eq.~\eqref{eq:dEdfsbare}, the frequency spectrum explicitly writes 
 \begin{equation}\label{eq:dEdfs}
     \frac{\text{d} E_{\rm GW}}{\text{d}f_s}  = \frac{q_{\bullet}\left(\pi G_{\rm N} M_\bullet^2\right)^{\frac23}}{3 (1+q_{\bullet})^{\frac13}}\frac{ M_\bullet^{\frac13} f_s^{-\frac{1}{3}}}{1+\left(\frac{f_{\rm c}}{f_s}\right)^{\frac{11}3} e^{-\left(\frac{f_{\rm sol}}{f_s}\right)^{\frac23}} }  \ .
 \end{equation}
 Here, $f_{\rm c}$ and $f_{\rm sol}$ are respectively the frequencies associated with $r_{\rm c}$ and $r_{\rm sol}$  by using the standard Kepler's law, while the exponential dependence arises from the soliton density in Eq.~\eqref{eq:rho}. As one can see, in the absence of $\mathcal W_{\rm DF}$  the standard power-law $\text{d}E_{\rm GW}/\text{d}f_s\propto f_s^{-1/3}$ is recovered. In the presence of dynamical friction two possible situations arise: 
 \begin{itemize}
 \item $f_{\rm c}<f_{\rm sol}$: In this case the frequency spectrum recovers the standard power-law behaviour. This is basically due to the fact that $r_{\rm sol}$ is smaller than $r_{\rm c}$ and therefore the dynamical friction of ULDM is always negligible regardless of the value of $f_s$. The benchmark denoted as the blue dotted line  in Fig.~\ref{fig:SolitonDensity} well represents this scenario. 
  \item $f_{\rm sol}<f_{\rm c}$: In this case the frequency spectrum can substantially deviate  from the standard behavior if $f_s$ falls in between $f_{\rm sol}$ and $f_{\rm c}$. In such frequency band the exponential in Eq.~\eqref{eq:dEdfs} is basically 1 while $f_{\rm c}/f_s>1$. Hence, $\text{d}E_{\rm GW}/\text{d}f_s\propto f_s^{10/3}$ which exhibits  a completely different frequency dependence with respect to the standard SGWB from SMBHs mergers. For frequencies either larger than $f_c$ or smaller than $f_{\rm sol}$ the standard behavior is recovered since either $f_c/f_s$ or the exponential in the denominator of Eq.~\eqref{eq:dEdfs} are negligible. 
 \end{itemize}

\smallskip
Having at our disposal the GW energy for a single source, we have to combine the signal with the whole SMBH population in order to compute the strain in Eq.~\eqref{eq:strain}. As mentioned before, this can be  done by expressing the quantity $\text{d}n_s/\text{d}z \text{d}\mathbf{X}$ in terms of the relevant BH parameters. However, it is much more convenient to work with the parameters of the single galaxy that host the two mergers, since these are directly measured. In particular, defining  $q_{\bigstar}=M_{\bigstar, 2}/M_{\bigstar, 1}<1$ and $M_{\bigstar}\equiv M_{\bigstar, 1}$ as the heaviest galaxy stellar mass of the merging pair, one gets
%
\begin{equation}\label{eq:strainFull}
\begin{split}
     h_c^2(f)= & \frac{3 H_0^2}{2\pi^2 \rho_c f^2}\int \text{d}z\,\text{d}q_{\bigstar}\,\text{d}M_{\bigstar}\frac{\text{d} n_s}{\text{d}z \, \text{d}q_{\bigstar} \, \text{d}M_{\bigstar}} \\ 
     & \times \frac{f_s}{1+z}\frac{\text{d}E_{\rm GW}}{\text{d}f_s}\bigg|_{q_{\bigstar},\,M_{\bigstar}} \ .
\end{split}
\end{equation}
%
Further details about the computation of the differential density population of SMBHs are given in Appendix~\ref{app:diffBHdensity}. 

\smallskip
The \NANO\ collaboration has reported in Fig.~1 of~\cite{NANOGrav:2023hfp} the experimental data-set of the GW strain in the frequency band $f\approx (2-30)$ nHz. In particular the data-point at $\bar{f}=0.12$ yr$^{-1}=3.93$ nHz provides the measurement of the strain with the best accuracy. Hence,  we will quote such data as a reference point  as follows and we will just use it to derive the limit on ULDM as explained  in more details in Sec.~\ref{Sec:results}.

Combining all the ingredients mentioned so far, the \textit{differential GW strain squared} ${\rm d} h_c^2/{\rm d} M_{\bullet}$  and the  strain $h_c$ are plotted in Fig.~\ref{fig:strains}. The differential strain  is crucial to understand which part of the SMBH population mostly contributes to the GW spectrum and it writes  
\begin{equation}\label{eq:diffstrain}
\frac{{\rm d} h_c^2(f)}{{\rm d} M_{\bullet}} = \frac{{\rm d} h_c^2(f)}{{\rm d} M_{\bigstar}} \frac{{\rm d} M_{\bigstar}}{{\rm d} M_{\bullet}} \ ,
\end{equation}
where ${\rm d} M_{\bigstar}/{\rm d} M_{\bullet}$ can be obtained from Eq.~\eqref{eq:MBHvsMstar}. It is reported in the left-panel of Fig.~\ref{fig:strains} keeping fix the frequency to $\bar{f}$. The right-panel of Fig.~\ref{fig:strains} shows instead the GW strain $h_c$ as a function of the frequency $f$. The blue region individuates the \NANO\  frequency band, while the vertical dashed line refers to the frequency $\bar{f}$. In all plots three benchmark values of the ULDM particles $m$ are denoted as colored lines, while the benchmark BH-soliton mass ratios $\varepsilon$ are reported with different dashing.  The pure GW strains which only come from SMBH mergers are shown in green. As is apparent  the strain $h_c$ recovers the standard power-law behavior  with $\beta_{\rm GW}=-2/3$, and, from the differential strain, one can appreciate that the peak of the signal remarkably matches the result in the literature, $M_{\bullet}\approx 10^{8.5} M_{\odot}$ \cite{NANOGrav:2023hfp} at $f=\bar f$. 

\smallskip
Coming back to the situation in which a solitonic core is switched on, from the qualitative arguments given below Eq.~\eqref{eq:dEdfs} we expect that for large enough $m$, the critical radius exceeds the soliton core, so that we enter in the regime of pure GW emission. This feature is well represented by the appearance of a turning point in the curve of the differential strain, which starts to recover the standard pure GW behavior. Conversely, to make the soliton core comparable to the critical radius, smaller BH masses are needed, 
 Eq.~\eqref{eq:coreradius}, so that the turning point appears at smaller BH masses when $m$ increases. For example, in reference to the solid black curve of the left-panel, we need $M_{\bullet}\simeq 2\times 10^9 M_{\odot}$ in order to have $r_{\rm c}\approx r_{\rm sol}$, while for the solid red curve we need $M_{\bullet}\simeq 8 \times 10^{8} M_{\odot}$. On the other hand, the effects of very small ULDM masses are evident in the right-panel of Fig.~\ref{fig:strains}. As a matter of fact, smaller masses shift the deformation of the strain at lower frequencies since the critical frequency $f_c$ scales as $f_c \sim m^{36/22}$ (in virtue of the Kepler's law applied to Eq.~\eqref{eq:rc}) and dramatically decreases at smaller masses. 

\section{Results}\label{Sec:results}
\begin{figure}[t!]
\begin{center}
	\includegraphics[width=0.95\textwidth]{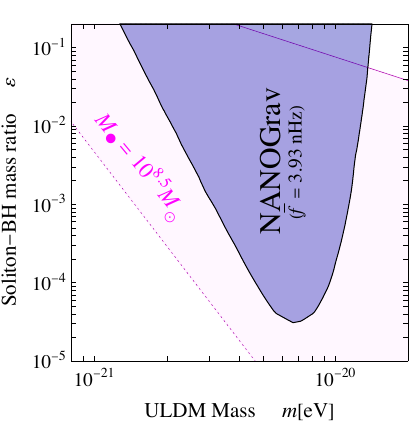}\hfill
	\caption{The \NANO\ bound in the $(m,\varepsilon)$ plane. The scaling relations suggested by DM-only numerical simulations and SMBH dominance are shown as magenta lines with different dashing (solid: DM-only numerical simulations, dotted: SMBH dominance).}\label{fig:results}
\end{center}
\end{figure}

 In order to derive limits on the properties of ULDM particles we compare the theoretical prediction of the GW strain given in Sec.~\ref{Sec:strain} with the experimental data by keeping  $f=\bar f$ fixed. This procedure allows us to set a constraint in the $(m,\varepsilon)$ plane by assessing the departure from the pure GW emission.  It is a conservative and robust approach because any potential additional astrophysical effects could only strengthen the constraints, albeit at the expense of robustness. The purple shaded region in Fig.~\ref{fig:results} denotes the parameter values for which the predicted strain exceeds the error bar. As one can see, the \NANO\ bound rules out BH-soliton mass ratios $\varepsilon< 0.2$ in the range of $m$ between $1.3\times 10^{-21} \, \text{eV}$ to $1.4\times 10^{-20} \, \text{eV}$. This can be understood in terms of the soliton properties.  Indeed, from Eq.~\eqref{eq:coreradius} one can easily read that the soliton radius scales as $m^{-2}$, while the critical distance in Eq.~\eqref{eq:rc} scales  as $r_{\rm c}\sim m^{-12/11}$. As a consequence, for $m\gtrsim 10^{-20}$ eV the critical radius exceeds the soliton core. Hence, the GW strain recovers the standard power-law behavior. As already commented in Sec.~\ref{Sec:strain} this happened at a BH mass $M_\bullet \simeq 8\times 10^8 M_\odot$, when the red lines of the differential strain in Fig.~\ref{fig:strains} dramatically change their behavior. On the other hand, for $m\lesssim 10^{-21}$ eV, the critical radius increases, the deformation of the GW strain is shifted at lower frequency (as one can see in the right-panel of Fig.~\ref{fig:strains}) and it goes outside the NANOGrav frequency band. The best sensitivity to ULDM is provided at $m\simeq 7\times 10^{-21}$ eV, when $\varepsilon\simeq 3 \times 10^{-5}$.
 
In addition to the NANOGrav bound, we also compare such limit with two classes of complementary constraints. The former  solely refer to the soliton properties and are given by:

\begin{itemize}
 \item The life-time of the soliton in presence of SMBHs: As discussed in Sec.~\ref{Sec:abstime}, the depletion of the soliton occurs in a characteristic time given by Eq.~\eqref{eq:abstime}. This must be compared to the age of the universe when the bulk of the merging events occurred ($z\approx 0.3$,~\cite{NANOGrav:2023hfp, Dror:2021wrl}), corresponding to $\tau_{\rm U}\approx 10.3$~Gyr. By choosing the BH peak mass coming from standard SGWB signal, $M_{\bullet}\approx 10^{8.5} M_\odot$, solitons formed by the condensation of ULDM particles with $m\gtrsim 10^{-21}$ eV are absorbed. Nevertheless, we recall that this limit is subject of large uncertainties coming from two main reasons: $i)$ the absorption time strongly depends on the BH and point particle masses. In particular, since for slight increases of $m$ the BH peak mass in the left-panel of Fig.~\ref{fig:strains} significantly shifts at lower value, the life-time of the surrounding solitons can span several orders of magnitude. For example, choosing a peak mass of $2\times 10^7 M_\odot$ (black solid line in the left-panel of Fig.~\ref{fig:strains}) the bound on the ULDM mass from soliton absorption becomes  $m\gtrsim 10^{-20}$ eV; $ii)$ the computations which lead to Eq.~\eqref{eq:abstime} do not account for the effects related to the gravitational backreaction of  the surrounding ULDM halo and the possible re-condensation of the soliton.  As a consequence we do not show this limit for lack of robustness. 
 
 When considering other processes that could potentially deplete the soliton, it is important to mention the tidal deformation of the core induced by the companion SMBH, as discussed in~\cite{Cardoso:2020hca}. Nevertheless, these effects do not significantly impact the parameter space of our analysis due to the relatively small gravitational coupling constant. Specifically, for values of $A$ on the order of $10^{-2}$, the tidal field introduced in Ref.~\cite{Cardoso:2020hca} is several orders of magnitude below the threshold for weak tidal deformation.

    \item The mass of the soliton: It is interesting to compare the observational constraints
of Fig.~\ref{fig:results} to theoretical expectations for the soliton mass. In Sec.~\ref{Sec:SolMass} we have discussed the scaling relations for the soliton mass given in Eq.~\eqref{eq:Msol_sim} and Eq.~\eqref{eq:Msol_BH} which represent  the result of numerical simulations and the naive estimate obtained in the full SMBH dominance respectively.  In particular, we focus on masses $m\gtrsim 10^{-21}$ eV and consider  $M_h = 5\times 10^{12} M_{\odot}$ and $M_{\bullet}= 10^{8.5} M_{\odot}$ as reference values for the galactic halos and SMBH peak masses.  In Fig.~\ref{fig:results} the scaling relations suggested by DM-only numerical simulations and SMBH dominance are shown as magenta lines with different dashing (solid: DM-only numerical simulations, dotted: SMBH dominance).

\end{itemize}

The second class of complementary bounds is related to the possibility that ultralight scalar fields fullfilled the total DM abundance. In such scenario, the  wave-like behaviour of ULDM can leave an imprint in the early Universe affecting astrophysical observations. In particular, the most relevant is given by the latest Lyman-$\alpha$ (Ly-$\alpha$) limit which rules out ULDM masses below $2\times10^{-20} \, \text{eV}$~\cite{Rogers:2020ltq, Irsic:2017yje, Armengaud:2017nkf, Zhang:2017chj}. Furthermore, a recent study on a group of ultra-faint dwarf galaxies also reports a tension with ULDM lighter than  $10^{-21} \, \text{eV}$~\cite{Hayashi:2021xxu}. However, as briefly stressed above, we have to keep well in mind that the bound derived in~\cite{Rogers:2020ltq} relies on the assumption that DM is only composed by ULDM particles. If ULDM only contributes to a subdominant fraction of a bigger Dark Sector, the Ly-$\alpha$ constraint  can be significantly weakened. Hence we do not show such limit in Fig.~\ref{fig:results}. 

\section{Conclusions and outlook}\label{Sec:conclusions}

In this paper, we have studied the possibility of using the recent \NANO\ SGWB measurements in the nHz frequency band to put constraints on ULDM.

\smallskip
These candidates, with typical masses in the $10^{-22} \, \text{eV} \lesssim m \lesssim 10^{-20} \, \text{eV}$ range, give rise to dense cores in the galactic centers, called solitons, which affect the merging of the SMBH binaries during the GW emission. Indeed, we have shown that the dynamical friction of the SMBHs with the solitons dominates over the effects sourced by the surrounding environment (stellar scattering, baryon dynamical friction and viscous drag~\cite{Kelley:2016gse}) and dramatically reduces the kinetic energy of the rotating SMBHs, so that the characteristic GW strain $h_c(f)$ is softened with respect to the one of  a binary system dominated by pure GW emission. 

\smallskip
On a more specific level, we have presented our main results in Figs.~\ref{fig:strains} and~\ref{fig:results}. In Fig.~\ref{fig:strains} we showed the effect of the soliton dynamical friction on the strain for different values of $\varepsilon= M_\text{sol}/M_\text{BH}$ to enlighten the softening of the signal in the nHz band. In addition to the full strain we also have reported the differential strain with respect to the BH mass, thus inspecting which part of the SMBH population contributes the most to the GW signal. In the $\varepsilon=0$ limit we have recovered a well known result: SMBHs with masses around $10^{8.5} M_{\odot}$ mostly contribute to the characteristic strain at the benchmark frequency  $\bar f = 3.93$ nHz. On the other hand, the novel effect of the soliton is to shift the peak of the strain at lower SMBH masses. This feature is fundamental when considering the depletion of the soliton due the SMBH, since low BH masses favor absorption times that are comparable to the age of the universe. In Fig.~\ref{fig:results} we have discussed how the dynamical friction sourced by the soliton affects the \NANO\ measurements by showing which parameter values of the $(m,\varepsilon)$ plane modify the strain significantly at $\bar f$. The excluded region runs in the  $1.3\times10^{-21} \, \text{eV} \lesssim m \lesssim 1.4\times10^{-20} \, \text{eV}$ range for $10^{-5}\lesssim\varepsilon\lesssim 0.2$. 

\smallskip
DM-only numerical simulations find a scaling relation that predicts $M_{\rm sol}$ in terms of the host halo and  ULDM masses. The magenta solid line, which is derived from the aforementioned naive extrapolation of the scaling relation,  is completely covered by the \NANO\ excluded region for the whole ULDM mass range. Nevertheless, a number of theoretical considerations discussed in Section~\ref{Sec:ULDM} prompt us to question the adequacy of this naive extrapolation of the soliton-halo relation, as it could lead to a substantial overestimation of the soliton mass. For this reason, we have introduced a lower bound on $M_{\rm sol}$, represented by a dotted magenta line. This lower bound takes into account potential complications related to the formation of SMBHs that might inhibit the development of solitons. Despite these uncertainties, the \NANO\ constraint remains sufficiently stringent, spanning multiple orders of magnitude in the parameter space. As a result, even with the theoretical caveats surrounding the determination of soliton mass, our findings still yield a robust constraint on ultralight dark matter with minimal coupling to gravity.

\smallskip
Upcoming PTA experiments, such as \IPTA~\cite{Kaiser:2020tlg}, will explore frequencies up to around $\mathcal O(100)$ nHz and study a supermassive black hole (SMBH) population down to masses of approximately $M_{\bullet}\simeq 10^{6} M_{\odot}$~\cite{Ellis:2023owy}. This progress offers the potential to constrain significantly longer-lived solitons formed from the condensation of ULDM particles with larger masses compared to those currently probed by \NANO.

\acknowledgments
This work receives partial funding from the European Union---Next generation EU (through Progetti di Ricerca di Interesse Nazionale (PRIN) Grant No. 202289JEW4). We thank Daniele Gaggero for useful discussions.

\appendix
\section{Differential density population}\label{app:diffBHdensity}
The differential density population in Eq.~\eqref{eq:strainFull} can be conveniently expressed in terms of the galaxy-stellar mass functions (GSMF, $\Psi$), the merger rate of the galaxies ($\tau$) and a galaxy pair fraction (GPF, $P$)~\cite{Chen:2018znx}
\begin{equation}\label{pop}
    \frac{\text{d} n_s}{\text{d}z \,\text{d}q_{\bigstar}\, \text{d}M_{\bigstar}}= \frac{\Psi (M_{\rm Gal},z)}{M_{\bigstar}\log{10}}\frac{P(M_{\bigstar},q_{\bigstar},z)}{\tau(M_{\bigstar},q_{\bigstar},z)}\frac{\text{d}t}{\text{d}z}.
\end{equation}

 Here we have $\Psi=\text{d}n_s/\text{d} \log_{10}{M_{\bigstar}}$ and $P=\text{d}\mathcal{F}/\text{d}q_\bigstar$, $\mathcal{F}$ being the fraction of merging galaxy pairs. All these quantities can be written by using analytical expressions that involve a number of fitting parameters~\cite{Chen:2018znx}. In our analysis we employ the parameters provided in Ref.~\cite{NANOGrav:2023hfp} following the \textbf{GWOnly + Uniform} setup. Notice that the factor $\text{d}t/\text{d}z$ converts the rate with respect to time into a rate with respect to redshift and it is given by standard cosmology.

 An important point concerns the so called \textit{SMBH–Host relation}. Indeed, the emitted energy in GWs, Eq.~\eqref{eq:dEdfs}, is written in terms of the BH relevant parameters, while the integral in Eq.~\eqref{eq:strainFull} runs over the host galaxies parameters. We can write Eq.~\eqref{eq:strainFull} as a function of the host parameters by using a one-to-one correspondence between the galactic stellar bulge and the SMBH masses~\cite{Marconi:2003hj}
 
 \begin{align}\label{eq:MBHvsMstar}
     \log_{10}{\left(\frac{M_{\bullet}}{M_{\odot}}\right)}=\mu+\alpha_{\mu}\log_{10}{\left(\frac{M_{\rm bulge}}{10^{11} M_{\odot}}\right)}+\mathcal{N}(0,\epsilon_{\mu}).
 \end{align}
 Here $M_{\bullet}$ and $M_{\rm bulge}$ denote the BH and the stellar bulge masses respectively, while $\mathcal{N}$ is the normal distribution with mean $0$ and standard deviation $\epsilon_{\mu}$. As before, we take the fiducial values of the $\mu$, $\alpha_{\mu}$ and $\epsilon_{\mu}$ parameters from Ref.~\cite{NANOGrav:2023hfp}. Finally, $\log_{10} q_\bullet = {\alpha_\mu} \log_{10} q_\bigstar$ assuming that the normal distribution in the SMBH-Host relation gives a negligible contribution due to the small standard deviation $\epsilon_{\mu}$, as showed in~\cite{NANOGrav:2023hfp} and  $M_{\rm bulge}$ is related to the galaxy stellar mass as $M_{\rm bulge}= f_{\rm bulge} \cdot M_{\bigstar}$, where $f_{\rm bulge}\approx 0.615$ is the bulge fraction~\cite{NANOGrav:2023hfp}.  

\bibliographystyle{utphys.bst}
\bibliography{Biblio}

\end{document}